\documentclass{mn2e}
\usepackage{color}
\usepackage{graphicx}
\usepackage{dcolumn}
\usepackage{bm}

\begin{document}

\title[MGC: SMBH mass function]
{The Millennium Galaxy Catalogue: 
The local supermassive black hole mass function 
in early- and late-type galaxies}

\author[Graham et al.]
{Alister W.\ Graham$^{1}$\thanks{AGraham@astro.swin.edu.au}, 
Simon P.\ Driver$^{2}$, Paul D.\ Allen$^{1,2}$ and Jochen Liske$^3$\\
$^1$Centre for Astrophysics and Supercomputing, Swinburne University
of Technology, Hawthorn, Victoria 3122, Australia\\ 
$^2$SUPA\thanks{Scottish Universities Physics Alliance (SUPA)}, 
School of Physics \& Astronomy, University of St Andrews, 
North Haugh, St Andrews, Fife, KY16 9SS, UK\\
$^3$European Southern Observatory, Karl-Schwarzschild-Str. 2, 85748 Garching, Germany.
}

\date{Received 2006 Jan 01; Accepted 2006 December 31}
\pubyear{2006} \volume{000}
\pagerange{\pageref{firstpage}--\pageref{lastpage}}

\input{psfig}

\maketitle
\label{firstpage}

\begin{abstract}
  
We provide a new estimate of the local supermassive black hole mass function
using {\it (i)} the empirical relation between supermassive black hole mass
and the S\'ersic index of the host spheroidal stellar system and {\it (ii)}
the measured (spheroid) S\'ersic indices drawn from 10k galaxies in the
Millennium Galaxy Catalogue.  
The observational simplicity of our approach, and the direct measurements of
the black hole predictor quantity, i.e.\ the S\'ersic index, for both
elliptical galaxies and the bulges of disc galaxies makes it straightforward
to estimate accurate black hole masses in early- and late-type galaxies alike.
%
%
We have parameterised the supermassive black hole mass function with a
Schechter function and find, at the low-mass end, a logarithmic slope
($1+\alpha$) of $\sim0.7$ for the full galaxy sample and $\sim$1.0 for the
early-type galaxy sample.  Considering spheroidal stellar systems brighter
than $M_B = -18$ mag, and integrating down to black hole masses of $10^6
M_{\odot}$, we find that the local mass density of supermassive black holes in
early-type galaxies $\rho_{\rm bh, early-type} = (3.5\pm 1.2) \times 10^5
h^3_{70} M_{\odot}$ Mpc$^{-3}$, and in late-type galaxies $\rho_{\rm bh,
  late-type} = (1.0\pm 0.5) \times 10^5 h^3_{70} M_{\odot}$ Mpc$^{-3}$.  The
uncertainties are derived from Monte Carlo simulations which include
uncertainties in the $M_{\rm bh}$--$n$ relation, the catalogue of S\'ersic
indices, the galaxy weights and Malmquist bias.  The combined, cosmological,
supermassive black hole mass density is thus $\Omega_{\rm bh, total} = (3.2
\pm 1.2)\times 10^{-6} h_{70}$.  That is, using a new and independent method,
we conclude that $(0.007\pm0.003)h_{70}^3$ per cent of the universe's baryons
are presently locked up in supermassive black holes at the centres of galaxies.

\end{abstract}


\section{Introduction}

Two purely photometric properties of galaxies, or rather their
spheroidal\footnote{By the term `spheroidal', we mean an entire elliptical
  galaxy or the dynamically hot component of a disc galaxy.}  components, are
known to correlate strongly with a galaxy's supermassive black hole (SMBH)
mass $M_{\rm bh}$.
The first property is optical luminosity (Kormendy 1993; 
Franceschini et al.\ 1998; Magorrian et al.\ 1998).  Due to the observation
that SMBHs are associated with the `bulge' of a galaxy, and not the disc, it
is necessary to perform a bulge/disc decomposition if one is to properly treat
lenticular and late-type galaxies.  At present, apart from (Erwin, Graham \&
Caon 2002; their figure~3, with only eight elliptical and five disc galaxies),
no optically\footnote{A useful near-infrared $M_{\rm bh}$--$L_{\rm spheroid}$
  relation is presented by Marconi \& Hunt (2003), updated in Graham (2007).}
calibrated relation that pertains to both elliptical galaxies and the bulges
of disc galaxies is available.
%
%
On the other hand, one can simply exclude the disc+bulge galaxies and only
work with elliptical galaxies (e.g., McLure \& Dunlop 2002, 2004).
%

In an effort to include disc galaxies, some authors have assigned some fixed
fraction, such as three-tenths, of a disc galaxy's total light to that of the
bulge.
However, given there is a known trend of decreasing bulge-to-total luminosity
ratio with increasing morphological type (e.g., Hubble 1926, 1936; 
Kent 1985; Simien \& de Vaucouleurs 1986; Andredakis,
Peletier \& Balcells 1995), the above approach introduces a systematic bias
such that the SMBH masses are over-estimated in the late-type disc galaxies
and under-estimated in the early-type disc galaxies, skewing the mid- to
low-mass end of the SMBH mass function.
Another approach has been to use the average bulge-to-total flux ratios
derived from past $R^{1/4}$-bulge $+$ exponential-disc decompositions 
(Simien \& de Vaucouleurs 1986).
However, Andredakis \& Sanders (1994) showed that Sb and Sc galaxies are, on
average, better described with an exponential-bulge than an $R^{1/4}$-bulge.
Andredakis et al.\ (1995) subsequently showed that an $R^{1/n}$-bulge was
more appropriate, with the S\'ersic index $n$ shown to decrease with
increasing disc galaxy type.
While an $R^{1/4}$-bulge plus exponential-disc results in an over-estimate of
the $B/D$ ratio when the bulge has a S\'ersic index $n<4$ (c.f. Figures~15 and
1 in Graham 2001), the use of an exponential-bulge $+$ exponential-disc
decomposition results in an under-estimate of the bulge-to-disc ($B/D$) flux
ratio when the bulge actually possesses a S\'ersic profile with $n>1$ (Graham
2001, his figure~13).  Therefore, $R^{1/n}$-bulge plus exponential-disc fits
are required.  
The various $M_{\rm bh}$-$L$ relations in the literature predict SMBH masses
that differ by factors of two to ten depending on the luminosity.  This
obviously inhibits the use of the $M_{\rm bh}$-$L$ relation at present. 
An investigation of this problem is not undertaken here but presented 
in Graham (2007).

The second\footnote{A third photometric quantity that has been {\it predicted}
  to correlate well with SMBH mass is the central stellar density of the host
  spheroid (Graham \& Driver 2007a, their section~6).}  photometric quantity
known to correlate with $M_{\rm bh}$ is the concentration of the stars in the
host spheroidal stellar system (Graham et al.\ 2001, 2003).  This concentration
is monotonically related to the shape, i.e.\ the S\'ersic index $n$, of the
spheroid's light-profile (Trujillo, Graham \& Caon 2001, their equation~6).
Moreover, the $M_{\rm bh}$--$n$ relation is known to be as tightly correlated
as the $M_{\rm bh}$--$\sigma$ relation and have the same small degree of
scatter 
(see Novak, Faber \& Dekel 2005 for a recent comparison of these relations).

\begin{figure}
\includegraphics[angle=270,scale=0.63]{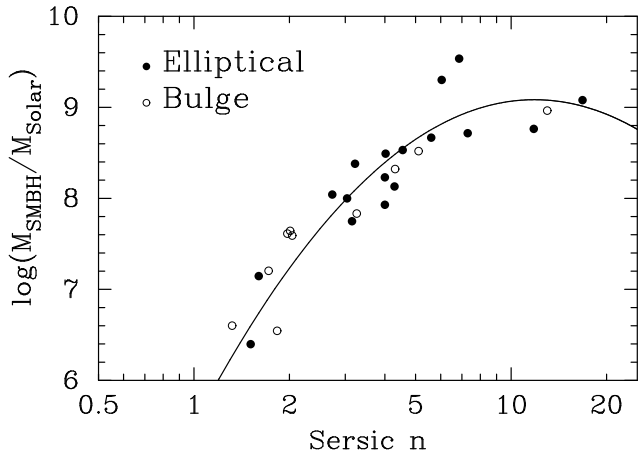}
\caption{
Relationship between SMBH mass and host spheroid S\'ersic index. 
The data points were taken from Graham \& Driver (2007a, their table 1) 
and the curve shows the best quadratic fit (equation 1). 
\label{MGC-m-n}
}
\end{figure}

Using an expanded galaxy set (27 galaxies) with updated 
distances and black
hole masses, Graham \& Driver (2007a) have recently shown
that the $M_{\rm bh}$--$n$ relation (see Fig.\ref{MGC-m-n}) is
curved rather than linear.  Fitting a quadratic equation, they
obtained
\begin{eqnarray}
\log (M_{\rm bh}) & = & 7.98(\pm0.09) +  3.70(\pm 0.46)\log(n/3) \nonumber \\
& & -3.10(\pm 0.84)[\log(n/3)]^2,
\label{EqQuad}
\end{eqnarray}
with an intrinsic scatter $\epsilon_{\rm intrinsic} = 0.18^{+0.07}_{-0.06}$ dex.
The total absolute scatter in $\log M_{\rm bh}$ is 0.31 dex, 
which compares favourably with the value of 0.34 dex from 
the $\log M_{\rm bh}$-$\sigma$ data and relation in Tremaine et al.\ (2002).
The parameter in front of the second order term in Eq.\ref{EqQuad} is 
inconsistent with a value of zero at the 99.99 per cent 
confidence level. 
%
%

\begin{figure*}
\includegraphics[angle=270,scale=0.63]{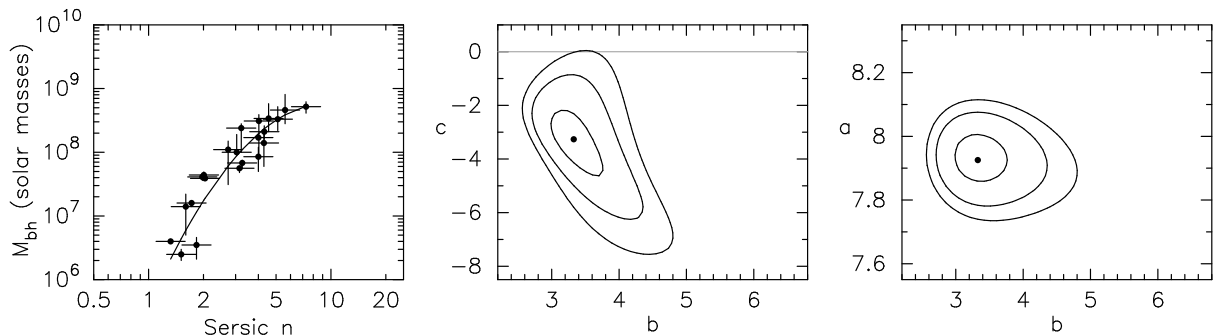}
\caption{Left panel: 
Variant of Fig.\ref{MGC-m-n} after removing the five data points 
with the highest SMBH masses.   The curved line 
corresponds to $\log M_{\rm bh}  =  a +  b\log(n/3) + c[\log(n/3)]^2$. 
Projections of the $\Delta \chi^2=1.00, 4.00$ and 6.63 contours, 
shown in the central and right panels, 
onto each axis gives the 68.3, 95.4, and 99 per cent confidence 
interval on each of the parameters $a, b$ and $c$. 
The optimal log-quadratic relation (shown in the left panel) 
is consistent with zero intrinsic scatter and reveals that the
coefficient, $c$, in front of the quadratic term is inconsistent with a value of
zero at the $\sim 2.5$ $\sigma$ level.  Subsequently, one can conclude that the 
log-quadratic relation in Eq.\ref{EqQuad} is not the result of increased
scatter, i.e.\ outliers, at the high-mass end of the $M_{\rm bh}$--$n$ 
relation (see Graham \& Driver 2007a for more details). 
\label{M-n-5}
}
\end{figure*}

We have explored here whether the departure from a linear relation may have
been driven by an increased and uneven scatter, i.e.\ outliers\footnote{As
  noted in Graham \& Driver (2007a, their Section~3.3), there is reason to
  suspect that the highest SMBH mass, pertaining to NGC~4486, may have been
  overestimated, perhaps by a factor of 4.}, at the high-mass end of the
$M_{\rm bh}$--$n$ relation, or whether the curvature is inherent in the rest
of the data set.  In Fig.\ref{M-n-5} we show the results of fitting a
log-quadratic relation after the removal of the five highest mass data points
from Fig.\ref{MGC-m-n}.  The coefficient in front of the quadratic term is
again found to be inconsistent with a value of zero, this time at the 99 per
cent level, and all three coefficients remain consistent, at the 1$\sigma$
level, with the values given in Eq.\ref{EqQuad}.

Support for the $M_{\rm bh}$--$n$ relation stems from its application
to galaxies not included in its construction.  For example, 
De Francesco et al.\ (2007) have recently measured a S\'ersic index $n=4.1$ for NGC~3998, 
from which one would predict $\log(M_{\rm bh}/M_{\sun}) = 8.43$, in perfect agreement 
with the mass they derived from a kinematical study of the nuclear gas. 
In another example, 
Guhathakurta et al.\ (2006, their Figure~S2) report a S\'ersic index $n=2$ 
for M31, which has a SMBH mass equal to 3.5$\times10^7 M_{\sun}$ 
(Ferrarese \& Ford 2005) 
and is therefore also in good agreement with the data in Fig~\ref{MGC-m-n}.
Furthermore, for a sample of 11 narrow-line Seyfert galaxies (Ryan et al.\
2007), the $M_{\rm bh}$--$n$ relation predicts SMBH masses in agreement
with those derived using the size of the broad line region and the
continuum flux, and suggests a problem with the ($M_{\rm bh}$--luminosity)-derived masses. 

The $M_{\rm bh}$-$n$ relation also implies a maximum mass to which SMBHs have
formed.  The broad turn-over seen in Fig.\ref{MGC-m-n} peaks at $n=11.9$,
where the predicted 1$\sigma$ range of SMBH masses spans 0.8 to 3.8$\times 10
^9 M_{\sun}$ (Graham \& Driver 2007a, their Eq.8).  In Graham \& Driver (2005,
their figure~1), one can see that for $n >= 5$ there is not that much
difference in profile shape, and hence there should not be much difference in
SMBH mass for galaxies with $n >= 5$.  Above $n \sim 12$, increasing $n$ to
infinity has almost no effect on the profile shape, and hence the SMBH masses
should all be the same.  Some kind of asymptotic-like $M_{\rm bh}$--$n$
function is therefore in some sense demanded by the form of the S\'ersic
$R^{1/n}$ model.  At the high-$n$ end of the distribution, 
we do not believe that galaxies with $n > 11.9$
have smaller SMBH masses than those with $n < 11.9$, and we note that the
log-quadratic $M_{\rm bh}$--$n$ relation (held fixed for $n>11.9$) appears more
logical than say a rising linear $M_{\rm bh}$--$n$ relationship.  In passing, we
note that the highest SMBH mass which has been directly measured in a quasar 
using reverberation mapping is only 2.6$\times 10^9 M_{\sun}$ (S50836+71, 
Kaspi et al.\ 2007) and is thus consistent with our predicted upper 1-$\sigma$ range.
The second highest quasar SMBH mass is 0.9$\times 10^9 M_{\sun}$ (3C273). 

In this paper we employ the $M_{\rm bh}$--$n$ relation in Eq.\ref{EqQuad} 
to derive the SMBH mass function using data from the Millennium Galaxy Catalogue
(MGC), which is described in Section~2.
Preliminary results have been presented in Driver et al.\ (2006b, 2006c). 
%
%
Using the bulge-disc decompositions of the brightest $10^4$ MGC
galaxies (Allen et al.\ 2006), in Section~3 we construct the 
SMBH mass function.  In Section~4 we compare our results with previous
efforts to measure the SMBH mass density, $\rho_{\rm bh,0}$, using other means. 
A summary of our analysis is provided in Section~5.  

Throughout this paper, unless specified otherwise,
we use $\Omega_{\Lambda}=0.7, \Omega_{M}=0.3$ and 
$h_{70}=H_0/(70$ km s$^{-1}$ Mpc$^{-1})$.

\section{The Millennium Galaxy Catalogue, and our spheroid sample}

The MGC is a medium-deep ($B_{\rm MGC}=24$ mag)\footnote{$B_{\rm MGC}$ is the
  Galactic extinction corrected,
  SExtractor (BEST, Vega) apparent magnitude (Liske et al.\ 2003).}
imaging survey of the nearby universe with a median seeing of
1.27$^{\prime\prime}$
and it has 96.1 per cent complete (99.8 per cent for $B_{\rm MGC} < 19.2$ mag)
redshift information for the MGC-BRIGHT sample of 10,095 objects with 
$B_{\rm MGC} < 20$ mag (Driver et al.\ 2005).  The imaging data was acquired with
the 2.5 m Isaac Newton Telescope which surveyed 37.5 square degrees\footnote{The
  actual usable area of sky reduced to 30.88 square degrees after excluding the
  ``bad'' regions (around bright stars, diffraction spikes, CCD defects, CCD
  gaps, CCD edges, vignetted corner, etc.).}  
in a 35 arcmin wide strip along
the equatorial sky from 10$h$ to 14$h$ 50$^{\prime}$ (Liske et al.\ 2003).
Each field was observed for 750 seconds through a Kitt Peak National
Observatory $B$-band filter (4407 \AA).
The survey reaches a depth of $\mu_{\rm limit} = 26$ mag arcsec$^{-2}$, with
objects catalogued down to $B_{\rm MGC} = 24$ mag. 
For comparison, the Sloan Digital Sky Survey Data Release 5 (Adelman-McCarthy
et al.\ 2007) has a median ($r$-band) seeing of 1.4$^{\prime\prime}$
and an effective exposure time of 54.1 seconds per band, leading to a $g$-band
(4686\AA) magnitude limit of 22.2 mag (roughly on an AB system, and with $B-g$
roughly one-third of a mag, Blanton \& Roweis 2007).

The MGC redshift information has come from a number of sources, as previously
detailed in Driver et al.\ (2005, their Table~1).
%
%
The median redshift is 0.12 and sample selection effects are well understood
(Driver et al.\ 2005; Liske et al.\ 2006).

\begin{figure}
\includegraphics[angle=0,scale=0.43]{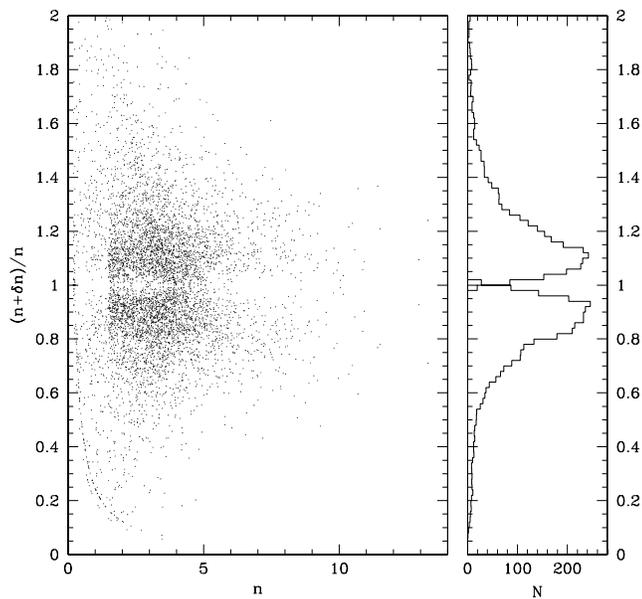}
\caption{
GIM2D-derived uncertainty in $n$, namely $\delta n$ (which need not be
symmetrical about $n$), is converted to a fractional uncertainty and
shown as a function of $n$ for the full MGC spheroid sample.  
Sixty eight per cent of the data has an error on $n$ of less than 20 per cent.
The bottom-left envelope of points in the left panel trace the lower limit
on the value of $n$ (equal to 0.2) that we set in GIM2D (see Allen et al.\ 2006).
A second envelope arises from the 0.09 dex colour-correction which is applied 
to the bulges of the late-type galaxies (see Section~\ref{Sec_colour}). 
\label{FigPaul}
}
\end{figure}

Allen et al.\ (2006) have performed an $R^{1/n}$-bulge plus exponential-disc
decomposition, using GIM2D (Marleau \& Simard 1998; Simard et al.\ 2002), for
all 10,095 objects.  In addition to the best-fitting bulge S\'ersic index $n$,
GIM2D derives the (not necessarily symmetric) upper and lower uncertainty,
$\delta n$, on the S\'ersic index.  In Fig.\ref{FigPaul} we show these
uncertainties in $n$ as a function of $n$.  The distribution is such that 68
per cent of the galaxies have an error on $n$ of less than $\sim$20 per cent.
Repeat observations, under different seeing conditions and on different 
chips of the wide-field camera, exist for 682 galaxies.  Fig.15 in Allen  
et al.\ (2006) shows the ability of GIM2D to consistently recover the 
S\'ersic index of the spheroid component.  The mean offset and standard     
deviation in the quantity $\Delta \log n$ from repeat observations was    
reported there to be $-$0.002 and 0.132 dex, respectively.  However, that 
distribution has longer tails than expected from a Gaussian.  This is due 
to 11 per cent of the objects whose galaxy `type' (see Allen et al.\ 2006)
was in disagreement.  Excluding these objects\footnote{The half-width of 
the 68 percentile of the full distribution
(i.e., all 682 galaxies) is 0.108 dex, which translates to a 28 per cent
mismatch in the value of $n$, slightly higher than the formal GIM2D
($1\sigma$) error.  This means that $\sim$11 per cent of the data may 
have a larger uncertainty than is assigned by GIM2D.}, 
the half-width of the 68   
percentile is 0.0729 dex, which corresponds to an 18 per cent mismatch in 
the value of $n$.  This figure agrees well with the formal GIM2D error
observed in Fig.\ref{FigPaul}, and also with the 20 per cent
uncertainty used in Graham \& Driver (2007a)
and commonly reported in the literature (e.g. MacArthur, Courteau \& Holtzman 2003). 
In this paper we adopted the GIM2D-derived values for both $n$ and $\delta n$,
as given in the catalogue 'mgc\_gim2d' which is publicly available at the MGC
website\footnote{http://www.eso.org/$\sim$jliske/mgc/}.
%
%

From the 10,095 galaxies in the MGC with $B_{\rm MGC} < 20$ mag, we
restrict the sample to those 7,745 objects with $0.013 < z < 0.18$.
Many of these are disc-only systems and therefore rejected as
potential black hole hosts.  
In passing we note that, once the attenuating effects of dust have been dealt
with, the MGC displays a uniformly flat distribution in $\cos(i)$,
where $i$ is the inclination of the disk, as one would expect for a 
uniformly distributed sample of galaxies (Driver et al.\ 2007b, their figure~5). 
We further refine our galaxy sample by imposing the requirement that both the
bulges and the elliptical galaxies, collectively referred to as `spheroids',
have half-light radii greater than 0.333 arcsec (1 pixel) and that the
bulge-to-total luminosity ratio ($B/T$) is greater than 0.01 (based on the
lower values observed by Graham 2001, his figure~15).  This helped avoid
bright nuclear components such as star clusters that may have been fitted with
the $R^{1/n}$ model in GIM2D.  We also required that the absolute spheroid
luminosity be brighter than $-18$ $B$-mag (discussed further in
Section~\ref{SecLow}).
Finally, we required that the galaxies core colour be red, such that
$(u-r)_{\rm core} > 2.00$ mag, denoting the transition in the colour
bimodality for the MGC (Driver et al.\ 2006a).  We refer to the red sample as
`Sample 3'.  However in Section~\ref{ColBias} we show that the effect of
including galaxies with blue cores (Sample 1 and 2) does not significantly
alter our results on the SMBH mass density.

\begin{figure}
\includegraphics[scale=0.43]{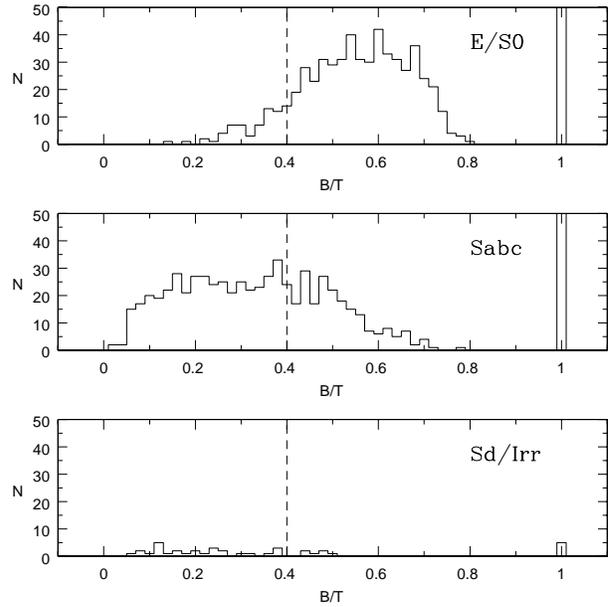}
\caption{Distribution of GIM2D-derived, bulge-to-total flux ratios 
($B/T$) for 
those MGC galaxies with an (eyeball) morphological classification 
(Driver et al.\ 2006a). 
Galaxies with a core colour bluer than $(u-r)_c=2.00$ mag have been 
excluded from the top panel. 
Spheroids fainter than $M_B = -17$ mag and effective half-light
radii smaller than 0.333 arcseconds (1 pixel) have been excluded from
all panels. 
The dashed line at $B/T=0.4$ denotes our adopted 
divide between early- and late-type galaxies.
\label{BT_Type}
}
\end{figure}

We additionally construct two mutually exclusive sub-samples, which we
label 'early-types' and 'late-types'. The distinction is based on
whether a galaxy's $B/T$ ratio is greater than or less than 0.4.
Our choice of 0.4 is lower than values of 0.5 or 0.6 which have often
been used in the past.  Figure~\ref{BT_Type} shows the $B/T$ ratio for
$\sim$3k MGC galaxies brighter than $B_{\rm MGC} = 19$ mag 
and which satisfy the above criteria and which
have also been classified by eye into three morphological bins (Driver et
al.\ 2006a): early-type galaxies (E/S0), early-type spiral galaxies
(Sabc) and late-type spiral galaxies (Sd/Irr).  If one was to use a
cut at $B/T=0.5$ or 0.6 to identify the early-type galaxies, then one
would miss roughly half of them.   
%
%

\subsection{Colour correction\label{Sec_colour}}

Before proceeding, we note that Eq.~\ref{EqQuad} was constructed
in the $R$-band, while the MGC galaxies have been imaged in the
$B$-band.  The presence of radial, $B-R$ colour gradients (e.g., La
Barbera et al.\ 2005, and references therein) may therefore result in
different values for the S\'ersic index $n$ in the two bands.

\begin{figure}
\includegraphics[scale=0.45]{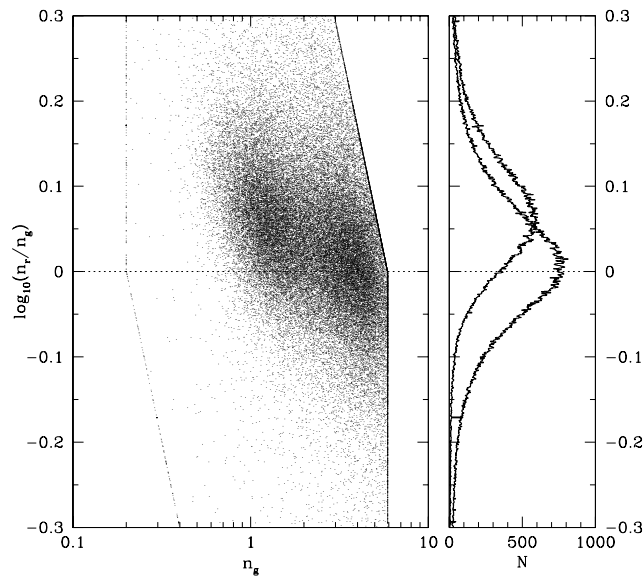}
\caption{Logarithmic difference between the $r$- and $g$-band S\'ersic
indices from individual galaxies in the SDSS-VAGC (Blanton et al.\
2005), in which a {\it single} $R^{1/n}$ function was fitted to each
galaxy and the S\'ersic index restricted to lie between 0.2 and 6.
The peak of the distribution for galaxies with $g$-band S\'ersic
indices greater than 2.0 (i.e., predominantly bulge-dominated,
early-type galaxies) is 0.003 dex.  The peak of the distribution for
galaxies with $n_g < 2.0$ (i.e., predominantly late-type galaxies) is
0.06 dex.
The choice of $n_g = 2.0$ originates from the S\'ersic index 
bimodality plots in Driver et al.\ (2006a).  
While the right-hand panel is based on all the SDSS-VAGC data, the
left hand panel shows a random sample of 75,000 with $n_g>2.0$ and
75,000 with $n_g<2.0$.  Such a restriction prevents washing away the
structure in the left panel with too many data points.
\label{SDSS-n}
}
\end{figure}

In general, colour gradients are known to be fairly small in
observations of local, early-type galaxies more luminous than 
$\sim -17$ $B$-mag (e.g., Peletier et al.\ 
1990; Taylor et al.\ 2005, and references therein).
The SDSS-VAGC (Blanton et al.\ 2005) has S\'ersic indices in the
$ugriz$ passbands in their low-redshift catalogue.  Although these
indices are somewhat different to ours, in that they are derived from
a single $R^{1/n}$-galaxy model rather than an $R^{1/n}$-bulge plus
exponential-disc model, we should be able to get some insight from
these data for the early-type galaxies, or at least the (disc-less)
elliptical galaxies.  An analysis of Fig.\ref{SDSS-n}, which shows
the difference between the SDSS $r$- and $g$-band S\'ersic index,
reveals that the median value of $\log(n_r/n_g)$ for galaxies with
$n_r > 2.0$ --- i.e.\ predominantly the early-type galaxies --- is
only 0.003 dex (after removal of the few obvious outliers with
absolute values greater than 0.3 dex).  We therefore apply no
correction to our $B$-band S\'ersic indices of the
early-type ($B/T > 0.4$) galaxies.

The single $R^{1/n}$-galaxy models that have been applied to the SDSS
data are not suitable for quantifying possible changes, with
wavelength, to the S\'ersic indices of bulges in late-type galaxies.
Instead, we use the $B$- and $R$-band S\'ersic indices from the bulges
of 86 disc galaxies given by Graham (2003), 79 of which have
indices in both bands.  The average ($\pm$ std.dev.) of the 79 values
of $\log(n_R/n_B)$ is 0.09 ($\pm0.15$).  Splitting the sample into
Sa--Sb and Sc--Sd--Sm galaxies gave the same small offset of 0.09 dex for
each grouping.  This suggests that the difference in $n$ is not a
function of spiral galaxy type nor bulge size.
In addition to radial stellar population gradients across the bulges, 
a plausible contributor to this difference is dust.  If the dust 
in spiral galaxies is more abundant at their centres, it will redden
their centres more than their outskirts, reducing the central 
parts of the $B$-band light profile relative to the $R$-band light profile and
thereby yielding smaller S\'ersic indices for the bulge in the $B$-band.
The similar $B$- and $R$-band S\'ersic indices for the early-type
galaxies suggests that dust is not a significant issue in these systems.
MacArthur et al.\ (2003) also provide $B$- and $R$-band S\'ersic indices
for an independent sample of 47 and 43 late-type spiral galaxies, respectively ---
with 42 in common.  In those instances where MacArthur et al.\ (2003)
provided multiple S\'ersic indices for the same galaxy in the same
passband, we averaged the logarithm of the S\'ersic indices. The
average ($\pm$ std.dev.) of the 42 values of $\log(n_R/n_B)$ is 0.08
($\pm0.17$), in good agreement with the data from Graham (2003).
In deriving the SMBH mass function using Eq.~\ref{EqQuad} (established
in the $R$-band), we will therefore apply a positive
correction of 0.09 dex to the logarithm of the $B$-band S\'ersic
indices from the bulges of our MGC late-type galaxies.

The presence of two distinct populations in Fig.\ref{SDSS-n}, rather than one
continuous distribution, suggests that the step-like correction we will apply
to the S\'ersic indices of our early- and late-type galaxies (0.0 dex and 0.09
dex respectively) may be more appropriate than a continuous correction based
on some parameter such as the $B/T$ ratio.

\section{SMBH mass function and space density}


To derive the SMBH mass function, we first modify the S\'ersic indices of the
late-type systems by 0.09 dex to convert from the $B$-band to the $R$-band, in
accord with the previous Section.  We then derive individual black hole masses
for each spheroid using equation~\ref{EqQuad}.  For each black hole we
determine an associated space-density weighting based on the MGC blue and red
spheroid luminosity functions as derived in Driver et al.\ (2007a). The weight
is the space density, $\phi(L)$, of the appropriate spheroid type (red or
blue, divided at $(u-r)_{\rm core} = 2.0$ mag) in the specified luminosity
interval, $L$, divided by the number of galaxies which contributed to that
interval, $N(L)$. The red and blue spheroid luminosity functions of Driver et
al.\ (2007a) were derived via `vanilla' step-wise maximum likelihood (see
Efstathiou, Ellis \& Peterson 1988) with K-corrections and e-corrections as
defined in Driver et al.\ (2007a).  The red and blue spheroid functions show
markedly different forms and the justification for the segregation into two
colour types is also given in Driver et al.\ (2007a).

\begin{figure}
\includegraphics[scale=0.45,angle=270]{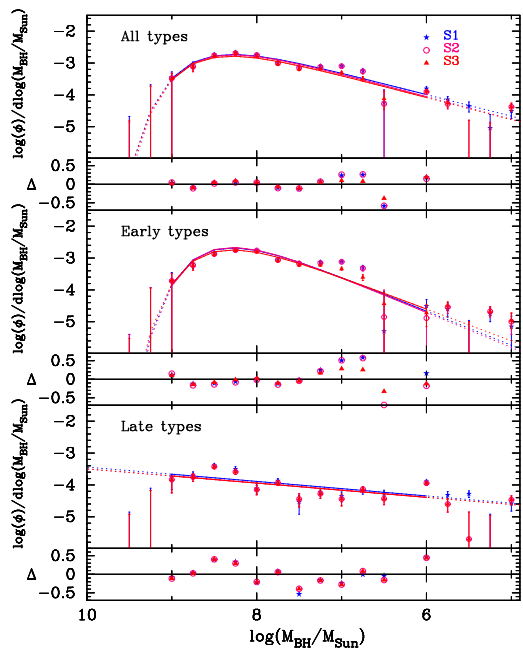}
\caption{ 
Black hole mass functions for our three (colour) samples, e.g.\ S3
= Sample 3, see Section~\ref{ColBias}. 
Early-type galaxies are those with $B/T>0.4$ (see Fig.\ref{BT_Type}).  
The best-fitting 3-parameter models 
(equation~\ref{Eqfit3}) are shown.  
Model parameters are given in Table~\ref{Tabfit}. 
The number density shown along the
vertical axis is expressed in units of $h^3_{70}$ Mpc$^{-3}$ per decade
in SMBH mass. 
The fitting is performed over the SMBH mass range $10^9 > M_{\rm
bh}/M_{\odot} > 10^6$.  Residuals about the models are shown beneath
each fit.  
\label{Figfit}
}
\end{figure}

The SMBH mass functions are then derived by summing the distribution of black
hole mass times weights, i.e., $\phi(M_{\rm bh})=\sum W(L)M_{\rm bh}$, where
$W(L)=\phi(L)/N(L)$.
This was constructed for black holes derived from all galaxies, early-types
only ($B/T > 0.4$) and late-types only ($B/T \leq 0.4$). 
See Fig.\ref{Figfit} for a graphic representations and
Table~\ref{massfn_table_sb} for a tabulated version of these distributions.

\begin{table}
\caption{
Supermassive black hole mass function data (corrected for Malmquist bias) for
the full, early- and late-type galaxy sample (Sample~3, see
Section~\ref{ColBias}) shown in Fig.~\ref{Figfit}. 
The uncertainties given are the upper and lower quartiles (i.e. $\pm$25 per cent) 
from extensive Monte Carlo realisation of the combined errors.} 
\label{massfn_table_sb}
\begin{tabular}{crrr} 
\hline
$\log_{10}$M$_{\rm bh}$ & \multicolumn{3}{c}{$\phi (10^{-4} h^3_{70}$ Mpc$^{-3}$ dex$^{-1})$} \\
$M_{\odot}$ & All galaxies & Early-type & Late-type \\ \hline
    5.00&    $0.42^{+0.13}_{-0.11}$ &    $0.11^{+0.08}_{-0.05}$ &   $0.34^{+0.11}_{-0.10}$ \\
    5.25&    $0.00^{+0.13}_{-0.00}$ &    $0.21^{+0.08}_{-0.06}$ &   $0.00^{+0.11}_{-0.00}$ \\
    5.50&    $0.00^{+0.16}_{-0.00}$ &    $0.00^{+0.10}_{-0.00}$ &   $0.02^{+0.12}_{-0.02}$ \\
    5.75&    $0.55^{+0.17}_{-0.16}$ &    $0.31^{+0.11}_{-0.09}$ &   $0.25^{+0.12}_{-0.11}$ \\
    6.00&    $1.29^{+0.20}_{-0.17}$ &    $0.19^{+0.16}_{-0.12}$ &   $1.13^{+0.13}_{-0.12}$ \\
    6.25&    $0.00^{+0.27}_{-0.00}$ &    $0.00^{+0.28}_{-0.00}$ &   $0.00^{+0.13}_{-0.00}$ \\
    6.50&    $0.76^{+0.51}_{-0.42}$ &    $0.37^{+0.47}_{-0.42}$ &   $0.37^{+0.14}_{-0.13}$ \\
    6.75&   $ 3.23^{+0.54}_{-0.54}$ &   $ 2.48^{+0.51}_{-0.52}$ &   $0.73^{+0.14}_{-0.13}$ \\
    7.00&   $ 4.97^{+0.54}_{-0.52}$ &   $ 4.61^{+0.54}_{-0.51}$ &   $0.36^{+0.15}_{-0.14}$ \\
    7.25&   $ 6.58^{+0.59}_{-0.61}$ &   $ 6.06^{+0.58}_{-0.58}$ &   $0.53^{+0.16}_{-0.15}$ \\
    7.50&   $ 6.40^{+0.72}_{-0.68}$ &   $ 6.02^{+0.67}_{-0.63}$ &   $0.36^{+0.17}_{-0.16}$ \\
    7.75&   $ 9.45^{+1.15}_{-1.01}$ &   $ 8.31^{+1.03}_{-0.90}$ &  $ 1.15^{+0.21}_{-0.19}$ \\
    8.00&   $16.29^{+1.74}_{-1.36}$ &   $15.55^{+1.54}_{-1.27}$ &  $ 0.71^{+0.26}_{-0.22}$ \\
    8.25&   $20.06^{+2.26}_{-1.89}$ &   $17.56^{+1.79}_{-1.50}$ &  $ 2.56^{+0.56}_{-0.42}$ \\
    8.50&   $16.79^{+1.85}_{-1.70}$ &   $13.02^{+1.57}_{-1.55}$ &  $ 3.73^{+0.49}_{-0.48}$ \\
    8.75&   $ 7.53^{+2.13}_{-2.22}$ &   $ 5.81^{+1.75}_{-1.71}$ &  $ 1.78^{+0.45}_{-0.49}$ \\
    9.00&    $3.26^{+1.41}_{-3.34}$ &    $1.96^{+1.31}_{-2.24}$ &   $1.47^{+0.33}_{-0.91}$ \\
    9.25&    $0.00^{+1.88}_{-0.00}$ &    $0.00^{+1.13}_{-0.00}$ &   $0.00^{+0.67}_{-0.00}$ \\
    9.50&    $0.00^{+0.16}_{-0.00}$ &    $0.00^{+0.03}_{-0.00}$ &   $0.00^{+0.12}_{-0.00}$ \\ 
\hline
\end{tabular} 

\noindent
Note: number densities are scaled to per unit $\log$M$_{\rm bh}$ interval and not per $0.25\log$M$_{\rm bh}$ interval.
\end{table}

There are a number of sources of potential error. We model these 
via Monte Carlo simulations, following both individually and collectively
the uncertainties on: the parameters 
defining equation~\ref{EqQuad}; the S\'ersic indices; the luminosity function
weights; and the Malmquist bias\footnote{Here we define Malmquist bias to be
  the systematic bias in our final measurements due to the presence of errors
  in our data.}. Errors not modelled at this stage because they are considered
secondary are uncertainties in the: K-corrections; e-corrections; dust
attenuation; spectroscopic incompleteness; photometric errors effecting the
magnitudes and hence weights; luminosity dependent cosmic variance; and the
choice of colour cut in the derived red and blue spheroid luminosity
functions.  Note that the global cosmic variance for the MGC survey area was
derived in Driver et al.\ (2005) by comparison to mock catalogues produced by
the Durham group\footnote{http://star-www.dur.ac.uk/$\sim$cole/mocks/};
this amounts to an overall uncertainty of 6 per cent 
in all $4 \times 10,001$ Monte Carlo simulations.

The Monte Carlo simulations consist of repeating the above analysis 10,001
times with the input values perturbed by the listed errors individually and
collectively, i.e., $4 \times 10,001$ Monte Carlo simulations in all.  All
error distributions are assumed to be Gaussian. The impact of the error(s)
is(are) then assessed by comparing the median and standard deviations of the
derived distributions to the original estimate.

Table~\ref{mgcrhos} summarises the cumulative sum of the SMBH mass functions,
$\rho_{\rm bh,0}$, with error estimates from each of the above sources. In all
cases we see that the dominant error is from the uncertainty in
equation~\ref{EqQuad}. This is perhaps not surprising since
equation~\ref{EqQuad} is based on a quadratic fit to only 27 systems for which
both credible black hole masses and S\'ersic indices exist. 
Improvements in this method will therefore come from 
a larger calibration sample (similarly for
the $M-\sigma$ and $M-L$ relations) and this should be attainable with next
generation facilities. Perhaps surprising is that the errors in the S\'ersic
distribution have relatively little impact. In part this is indicative of the
size of the sample but also helped by the quadratic nature of the $M-n$
relation preventing high $n$ values leading to unreasonably large $M_{\rm bh}$
values. The errors from the luminosity function (i.e., statistics and cosmic
variance) appear negligible, with cosmic variance the most significant.

\begin{table}
\begin{minipage}{85mm}
\caption{The space density of matter in supermassive black holes. 
The errors are 1-$\sigma$ values, after excluding 3-$\sigma$ outliers.
The densities given in
Table~\ref{massfn_table_sb}, rather than the fitted
empirical models, have been integrated down to SMBH
masses of $10^6 M_{\odot}$. 
The final column shows the density normalised against the critical 
density.  
The reason why the density varies
with $h^3$ and not $h^2$ is explained in Section~3.
Factoring in the intrinsic scatter, $\Delta$, from the $M_{\rm
bh}$--$n$ log-quadratic relation, the numbers in this Table should be
increased by $\exp[{(\Delta \ln 10)^2}/2]$, which equals 1.09 if
$\Delta=0.18$ dex (Graham \& Driver 2007a). 
\label{mgcrhos}
}
\begin{tabular}{@{}lccc@{}} 
\hline
\hline
          & No.      & 
$\rho_{\rm bh,0}$\footnote{Malmquist bias corrected.}  
$\pm\delta_1$\footnote{Monte Carlo simulation of the errors in the $M_{\rm bh}$--$n$ relation assuming a Gaussian error distribution.}
$\pm\delta_2$\footnote{Monte Carlo simulation of the error in $n$ as specified by GIM2D and assuming a Gaussian error distribution.}
$\pm\delta_3$\footnote{Monte Carlo simulations of the error in the individual galaxy weights assuming a Gaussian error distribution.}
$\pm\delta_4$\footnote{MGC global cosmic variance of 6 per cent for the effective $\sim 30$ sq degree region with $0.013 < z< 0.18$.} & 
$\Omega_{\rm bh}$ \\
                                               & Bulges   &  $h_{70}^3 10^{5} M_{\odot}$ Mpc$^{-3}$  &  $10^{-6}h_{70}$ \\
\hline 
\multicolumn{4}{c}{Early- and Late-type ($B/T>0.01$)} \\
Sample 1  & 1769     &  $4.87 \pm 1.84 \pm 0.07 \pm 0.01 \pm 0.29$  &  3.6$\pm$1.4  \\
Sample 2  & 1676     &  $4.53 \pm 1.63 \pm 0.07 \pm 0.01 \pm 0.27$  &  3.3$\pm$1.2  \\
Sample 3  & 1543     &  $4.41 \pm 1.65 \pm 0.07 \pm 0.01 \pm 0.26$  &  3.2$\pm$1.2  \\
[2pt]
\multicolumn{4}{c}{Early-type ($B/T > 0.4$)} \\
Sample 1  & 1539     &  $3.78 \pm 1.28 \pm 0.06 \pm 0.01 \pm 0.23$  &  2.8$\pm$1.0  \\
Sample 2  & 1485     &  $3.57 \pm 1.21 \pm 0.06 \pm 0.01 \pm 0.21$  &  2.6$\pm$0.9  \\
Sample 3  & 1352     &  $3.46 \pm 1.14 \pm 0.06 \pm 0.01 \pm 0.21$  &  2.5$\pm$0.9  \\
[2pt]
\multicolumn{4}{c}{Late-type ($0.01 < B/T \leq 0.4$)} \\
Sample 1  & 230      &  $1.14 \pm 0.56 \pm 0.04 \pm 0.01 \pm 0.07$  &  0.8$\pm$0.4  \\
Sample 2  & 191      &  $0.95 \pm 0.49 \pm 0.03 \pm 0.01 \pm 0.06$  &  0.7$\pm$0.4  \\
Sample 3  & 191      &  \multicolumn{2}{c}{as above}  \\
\hline
\end{tabular}
\end{minipage}
\end{table}

Finally, we correct for the Malmquist bias in a rather straightforward manner.
This is achieved by measuring $\rho_{\rm bh,0}$ assuming no errors and then
remeasuring it but allowing all of the above errors to be perturbed
simultaneously via Monte Carlo simulations. The difference between the two
estimated values (raw and the median of the Monte Carlo simulations) provides
a crude measurement of the systematic offset caused by the errors inherent in
our data. To provide Malmquist bias corrected values we then subtract this
systematic from our original values ($\sim 1$ to $3$ per cent downward
correction).  These final Malmquist bias corrected values are those shown in
Tables~\ref{massfn_table_sb} and \ref{mgcrhos}.

Fig.\ref{Figfit} shows the resulting black hole mass functions, where the
error bars indicate the combined errors derived from the Monte Carlo analysis
but excluding the cosmic variance error (as this is a hidden systematic and
not a random error). Our resulting SMBH mass functions (and number density and
mass density) depend on the Hubble constant.  As was shown in Graham \& Driver
(2007a), the $M_{\rm bh}$--$n$ relation is independent of the Hubble constant.
This is because the S\'ersic index $n$ does not depend on galaxy distance, and
while the SMBH mass does depend on distance, the overwhelming majority of
galaxies that were used to construct the $M_{\rm bh}$--$n$ relation had their
distances obtained by Tonry et al.\ (2001) using surface brightness
fluctuations --- a technique that provides distances without assuming some
Hubble constant.  The mass function dependence on the Hubble constant arises
from the $1/V_{\rm max}$ weighting given to each galaxy; it is this volume
term which introduces an $h^3$ dependence.

\subsection{Parameterisation of the SMBH mass function}

We have fitted two empirical models to our SMBH mass functions
over the mass range $10^6 < M_{\rm bh}/M_{\odot} < 10^9$.
The first is a mild variation of the commonly-used 3-parameter 
Schechter (1976) function, and is given by 
\begin{equation}
\phi(M_{\rm bh}) = \phi_* \left( \frac{M_{\rm bh}}{M_*} \right)^{\alpha+1}
{\rm exp}\left[ 1 - \left( \frac{M_{\rm bh}}{M_*} \right) \right], 
\label{Eqfit3} 
\end{equation}
where $\phi(M_*)=\phi_*$ (per unit ${\rm d} \log(M_{\rm bh})$ per Mpc$^3$). 
The turnover of the mass function and the 
maximum density occur at the SMBH mass
\begin{equation}
M_{\rm max} = (\alpha+1)M_* ,  \hskip20pt \alpha+1 > 0 \nonumber
\end{equation}
where the associated maximum density is 
\begin{equation}
\phi_{\rm max} = \phi_* (\alpha+1)^{\alpha+1} 
{\rm exp}\left[ - \alpha \right] . \nonumber
\end{equation}

The logarithmic slope at the low-mass end of the mass function 
is given by the exponent $1+\alpha$; a value of $\alpha=-1$ therefore
corresponds to a flat distribution, and larger values correspond to a 
decreasing function as the SMBH mass decreases. 
For the early-type galaxy samples the slope $(1+\alpha)$ 
is $\sim$1 (see Table~\ref{Tabfit}). 
For the (early+late)-type samples the slope is approximately two-thirds; 
the shallower decline is due to the contribution of SMBHs
in late-type galaxies. 
These, and the other, best-fitting parameters are reported in 
Table~\ref{Tabfit} and the fits themselves are shown in 
Fig.\ref{Figfit}.

\begin{table*}
 \begin{minipage}{110mm}
\caption{Best-fitting parameters from the empirical SMBH mass function
given in equation~\ref{Eqfit3}. 
The number density, $\phi_*$, is per decade in SMBH mass. 
The late-type galaxy sample is not shown here due to the poor match between 
the empirical model and the data (see Fig.\ref{Figfit}). 
Sample~2 excludes bulges in disc galaxies if the core colour
is bluer than $(u-r)_c = 2.00$ mag.  Sample~3 (our primary sample) excludes blue bulges and blue 
elliptical galaxies if $(u-r)_c \leq 2.00$ mag (see Section~\ref{ColBias}).
\label{Tabfit}
}
%
%
\begin{tabular}{@{}lccc@{}}
\hline
Data sample   &  $\log \phi_*$   &   $\log(M_*/M_{\odot})$   &   $\alpha$  \\
              &  $h^3_{70}$ Mpc$^{-3}$ dex$^{-1}$   &     &    \\
\hline
\hspace{3mm} Early- and Late-type, ($B/T > 0.01$): & \multicolumn{2}{c}{}  \\
Sample 1: no colour cut                             & -2.76  &  8.45  &  -0.32  \\
Sample 2: $(u-r)_{\rm c}{\rm [bulges]} > 2.00$ mag  & -2.76  &  8.43  &  -0.29  \\
Sample 3: $(u-r)_{\rm c}{\rm [all]} > 2.00$ mag     & -2.81  &  8.46  &  -0.30  \\ 
[2pt] 
\hspace{3mm} Early-type, ($B/T > 0.4$):  &  \multicolumn{2}{c}{}  \\
Sample 1: no colour cut                             & -2.68  &  8.26  &   0.07  \\
Sample 2: $(u-r)_{\rm c}{\rm [bulges]} > 2.00$ mag  & -2.67  &  8.25  &   0.10  \\ 
Sample 3: $(u-r)_{\rm c}{\rm [all]} > 2.00$ mag     & -2.74  &  8.29  &   0.00  \\
\hline
\end{tabular}
\end{minipage}
\end{table*}

We explored the suitability of a
second model having an additional fourth parameter
such that the mass term in the 
exponential of equation~\ref{Eqfit3} is raised to the power of $\beta$ to give 
\begin{equation}
\phi(M_{\rm bh}) = \phi_* \left( \frac{M_{\rm bh}}{M_*} \right)^{\alpha+1}
{\rm exp}\left[ 1 - \left( \frac{M_{\rm bh}}{M_*} \right)^{\beta} \right] 
\label{Eqfit4}
\end{equation}
(Aller \& Richstone, their equation~10).  
Our data, however, did not justify the 
need for this additional parameter.  For the (early+late)-type samples, the
value of $\beta$ equalled 1.0$\pm$0.1.  
For the early-type sample, it ranged from 0.3 to 0.5, but did not give 
rise to significantly better fits.  For this reason we do not show the 
fits or the parameters from this model. 
(Neither equations~\ref{Eqfit3} nor \ref{Eqfit4} 
provided an acceptable description to the 
SMBH mass function in our late-type galaxy sample.) 

%

\subsection{Integrating the SMBH mass function}

\subsubsection{SMBH number density}

The total SMBH number density can be obtained by 
integrating equation~\ref{Eqfit3} 
with respect to $\log(M_{\rm bh})$ (because $\phi$ is 
expressed in units of $h^3_{70}$ Mpc$^{-3}$ per {\it decade}
in SMBH mass). 
Integrating over $10^6 < M_{\rm bh}/M_{\odot} < 10^9$, 
the number density is given by the expression 
\begin{eqnarray}
 &&  \int_{\log(M_{\rm bh}/M_{\odot}) = 6}^{\log(M_{\rm bh}/M_{\odot}) = 9}
              \phi(M_{\rm bh}) \hskip3pt
              {\rm d}\log M_{\rm bh} \nonumber \\
 &=&  \int_{M_{\rm bh} = 10^6M_{\odot}}^{M_{\rm bh} = 10^9M_{\odot}}
              \frac{\phi(M_{\rm bh})}{\ln(10)M_{\rm bh}} \hskip3pt
              {\rm d} M_{\rm bh} \nonumber \\
 &=&  \frac{\phi_* \rm{e}^1}{\ln(10)} \left[ 
\gamma \left( \alpha+1,\frac{10^9 M_{\odot}}{M_*} \right) - 
\gamma \left( \alpha+1,\frac{10^6 M_{\odot}}{M_*} \right) \right], 
\end{eqnarray}
where $\gamma (a,x)$ is the incomplete gamma function 
(e.g., Press et al.\ 1992) defined by
\begin{equation}
\gamma (a,x)=\int ^{x}_{0} {\rm e}^{-t}t^{a-1}{\rm d}t.
\label{gamFunc}
\end{equation}
For the early- and late-type galaxy values of $\phi_*, M_*$ and
$\alpha$ (given in Table~\ref{Tabfit}), the number density 
of SMBHs with masses between one million and one billion solar masses is
$2.3\times10^{-3} h_{70}^3$ Mpc$^{-3}$.  
For the early-type galaxies, the number density is $2.1\times10^{-3} h_{70}^3$
Mpc$^{-3}$.

These values are an order of magnitude smaller than those reported in 
Shankar et al.\ (2004).  The difference can be attributed to their
rising ($M_{\rm bh}$--$L$)-derived 
SMBH mass function (for all galaxy types) as one moves towards lower-masses. 
In contrast, 
our ($M_{\rm bh}$--$n$)-derived mass function shows the opposite behaviour
at the low-mass end.

\subsubsection{SMBH mass density} 

In so far as equation~\ref{Eqfit3} represents the mass function
over the SMBH mass range $10^6 < M_{\rm bh}/M_{\odot} < 10^9$, the 
SMBH mass density for SMBHs having such masses can be obtained from 
\begin{eqnarray}
\rho_{\rm bh} = \int_{\log(M_{\rm bh}/M_{\odot}) = 6}^{\log(M_{\rm bh}/M_{\odot}) = 9}
              \phi(M_{\rm bh}) M_{\rm bh} \hskip3pt
              {\rm d}\log M_{\rm bh} \nonumber \\
  = \frac{\phi_* M_* \rm{e}^1}{\ln(10)} \left[ 
\gamma \left( \alpha+2,\frac{10^9 M_{\odot}}{M_*} \right) - 
\gamma \left( \alpha+2,\frac{10^6 M_{\odot}}{M_*} \right) \right]. 
\end{eqnarray}
Using the best-fitting parameters in Table~\ref{Tabfit}, we obtain 
$\rho_{\rm bh} = 4.3 \times 10^{5} h_{70}^3 M_{\odot}$ Mpc$^{-3}$ 
(all-types), and 
$\rho_{\rm bh} = 4.0 \times 10^{5} h_{70}^3 M_{\odot}$ Mpc$^{-3}$ 
(early-types). 
Integrating the mass function over {\it all} masses increases the above
values to $4.8 \times 10^{5} h_{70}^3 M_{\odot}$ Mpc$^{-3}$ and 
$4.2 \times 10^{5} h_{70}^3 M_{\odot}$ Mpc$^{-3}$, respectively.

We have, however, opted not to use the above formula, but to instead
acquire the SMBH mass densities directly from our data,
rather than from the fitted model.

Computing the local mass density, 
and integrating down to SMBH masses of $10^6 M_{\odot}$, we obtain 
$\rho_{\rm bh,early-type} = (3.5 \pm 1.2) \times 10^5 h_{70}^3 M_{\odot}$ Mpc$^{-3}$ 
while
$\rho_{\rm bh,late-type} = (1.0 \pm 0.5) \times 10^5 h_{70}^3 M_{\odot}$ Mpc$^{-3}$. 
These results are presented in Table~\ref{mgcrhos}. 

The above densities correspond to 
$\Omega_{\rm bh,early-type} = \rho_{\rm bh,early-type}/\rho_{\rm crit}
= (2.5 \pm 0.9)\times 10^{-6} h_{70}$, 
and
$\Omega_{\rm bh,late-type} = (0.7 \pm 0.4)\times 10^{-6} h_{70}$ (see Table~\ref{mgcrhos}), 
where $\rho_{\rm crit} = 3H_0^2/8\pi G$ is the critical density
for flat space\footnote{For $H_0=70$ km s$^{-1}$ Mpc$^{-1}$,  
$\rho_{\rm crit} \sim 1.36\times10^{11} M_{\odot}$ Mpc$^{-3}$.}.  
For reference, using an independent technique, 
Fukugita \& Peebles (2004, their equation~75)
give $2.5^{+2.5}_{-1.2}\times10^{-6}$ for the SMBHs in early-type galaxies, 
and  $1.3^{+1.2}_{-0.7}\times10^{-6}$ for the SMBHs in late-type galaxies.

%
In deriving the (linear) SMBH mass density, $\rho_{\rm
bh,0}$, (i.e., not $\log \rho_{\rm bh,0}$) from the convolution of the
distribution function of $n$ with the $\log M_{\rm bh}$--$\log n$
relation, one needs to allow for the intrinsic scatter in this $\log -
\log$ correlation.
If there is a Gaussian distribution of intrinsic scatter 
in the SMBH mass at any fixed $\log
n$, with a standard deviation which is independent of $\log n$ and equal to
$\Delta$, then the SMBH mass density should be increased by the factor
$\exp[{(\Delta \ln 10)^2}/2]$ (Yu \& Tremaine 2002, their
equation~12).
Such a Gaussian distribution, however, is not true for the scatter of
points about the log-quadratic $M_{\rm bh}$--$n$ relation shown in
Fig.\ref{MGC-m-n}; the scatter is clearly
greater at the high-mass end.  In fact, removal of the two highest
mass SMBHs results in a log-quadratic relation consistent with zero
intrinsic scatter (Graham \& Driver 2007a).  
It is thus questionable whether one should apply
this multiplicative term and because of this uncertainty, 
in what follows, we have not. 
However, one should note that if there were in fact no measurement
errors and $\Delta=0.31$ dex (the total absolute scatter about the
$\log M_{\rm bh}$--$\log n$ relation in Fig.\ref{MGC-m-n}), 
then the corrective factor to
apply to $\rho_{\rm bh,0}$ would be 1.29.  If $\Delta = 0.18$ dex
(the intrinsic scatter after factoring in the suspected measurement
errors), then the multiplicative factor drops to 1.09.  If there is no
intrinsic scatter, then the multiplicative factor is simply 1.
%

\subsubsection{SMBH baryon fraction}

The above numbers can be expressed in terms of the baryon fraction of the
universe.  Already a picture is emerging in which dark accretion onto
SMBHs may be negligible.  Shankar et al.\ (2004) claim that the local
SMBH mass function can be accounted for from mass accreted by X-ray
selected AGN, i.e.\ no significant `dark' accretion is required for
SMBH growth (see also Cao 2007).  If correct, this would imply that the growth of SMBHs from the
accretion of massive black hole remnants of Population III stars may
be small (Madau \& Rees 2001; Islam, Taylor \& Joseph 2003; Wyithe \&
Loeb 2004).  Moreover, baryonic-fuelling is consistent with a picture
that links the BH mass to the host spheroid's baryonic, or at least
stellar, properties, such as luminosity (e.g., Erwin, Graham \& Caon 2002;
McLure \& Dunlop 2002; Marconi \& Hunt 2003 and references therein), 
concentration (Graham et al.\ 2001; Graham \& Driver 2007a) 
and mass (Magorrian et al.\ 1998; H\"aring \& Rix 2004).
Assuming the above standard picture in which SMBHs form via the
accretion of baryons (Blandford 2004), then in terms of the invariant
baryon fraction of the total mass-energy density, such that
%
%
$\Omega_{\rm baryon}=0.0453h_{70}^{-2}$ (Tegmark et al.\ 2006, their Table~2,
row~3; see also Blake et al.\ 2007), 
{\it SMBHs at the centres of galaxies today contain 
$(0.007\pm0.003)h_{70}^3$ per cent of the universe's baryon inventory.}

Letting $M_{\rm spheroid}$ denote the stellar mass of a spheroid, 
H\"aring \& Rix (2004) have shown that 
$M_{\rm bh}/M_{\rm spheroid} = 0.0009$ when $M_{\rm bh}=10^6 M_{\odot}$
and $=0.0019$ when $M_{\rm bh}=10^9 M_{\odot}$, confirming 
the results given in Merritt \& Ferrarese (2001, their 
Section~3; see also Laor 2001). 
Taking the average (or using the peak in our SMBH mass function at
$\log(M_{\rm bh})=8.3$, see Fig.\ref{Figfit}) 
one obtains 
$M_{\rm bh}$ equals 0.14 per cent of $M_{\rm spheroid}$ (or 0.16 per cent). 
Dividing $\Omega_{\rm bh}/\Omega_{\rm baryon}$ by this value 
one obtains $\Omega_{\rm spheroid} \sim 6$ per cent (or $\sim$5 per cent) 
of $\Omega_{\rm baryon}$.  This is of course just a rough estimate, 
and a more precise value obtained using the actual spheroid luminosity 
function will be presented in Driver et al.\ (2007a, 2007b).

\subsubsection{Comments on the low mass end \label{SecLow}}

\begin{figure}
\includegraphics[scale=0.45]{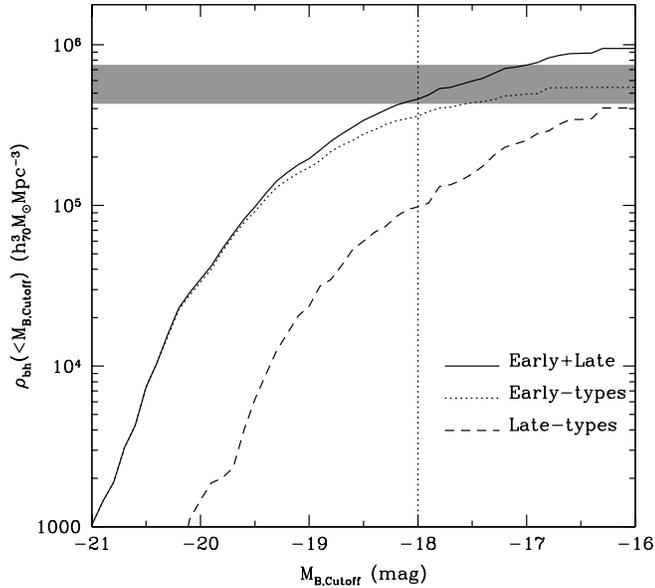}
\caption{Cumulative SMBH space density, $\rho_{\rm bh}$. 
Our early-type galaxies are those with $B/T > 0.4$, while 
our late-type galaxies have $0.01 < B/T \leq 0.4$. 
The cutoff at an absolute magnitude of $-$18 B-mag 
marks the boundary to which we can trust our data.
The horizontal shading is centred on the ($H_0$ corrected, i.e.\
41 per cent increased)  solution for (early+late)-type 
galaxies by Shankar et al.\ (2004, their section 3.2), 
and shows their quoted 1-$\sigma$ uncertainty. 
\label{Fig_rho}
}
\end{figure}

Figure~\ref{Fig_rho} shows how the SMBH space density is built up
as one includes (intrinsically) fainter spheroids, 
for our three galaxy bins.  
As was noted in Section~2, we excluded spheroids fainter
than $M_B = -18$ mag in our analysis.  
For the early-type galaxy sample, Figure~\ref{Fig_rho} reveals 
that integrating down to an absolute magnitude of $-16$ $B$-mag
gives a SMBH mass density that is consistent (at the 2-$\sigma$ level) with the 
result obtained using only those spheroids brighter than $-18$ $B$-mag.  We do 
however caution that the SMBH masses are less reliable in this
regime, and as such we do not wish to place too much emphasis on this
result. 

It is also worth noting that the central massive object in spheroids
fainter than $M_B \sim -18$ mag is often a nuclear star cluster 
(Ferrarese et al.\ 2006, left panel of their figure~2; Balcells et al.\ 2007).  
It is therefore questionable as to whether or not one should include such
faint spheroids in an analysis of this kind. 
Moreover, if they are not individually modelled, the presence of
additional nuclear star clusters can bias the S\'ersic $R^{1/n}$ fit
to give spuriously high values of the index $n$.  The effect is to
over-estimate the contribution from SMBHs in faint spheroids.  Given
the high frequency of nucleated bulges in spiral galaxies (e.g.,
Carollo, Stiavelli, \& Mack 1998; 
Balcells et al.\ 2003), there is thus reason to doubt the
rising SMBH space density curve shown in Fig.\ref{Fig_rho} for the
late-type galaxies.

\section{Comparison with past estimates} \label{SecComp}


%

\begin{figure}
\includegraphics[scale=0.77]{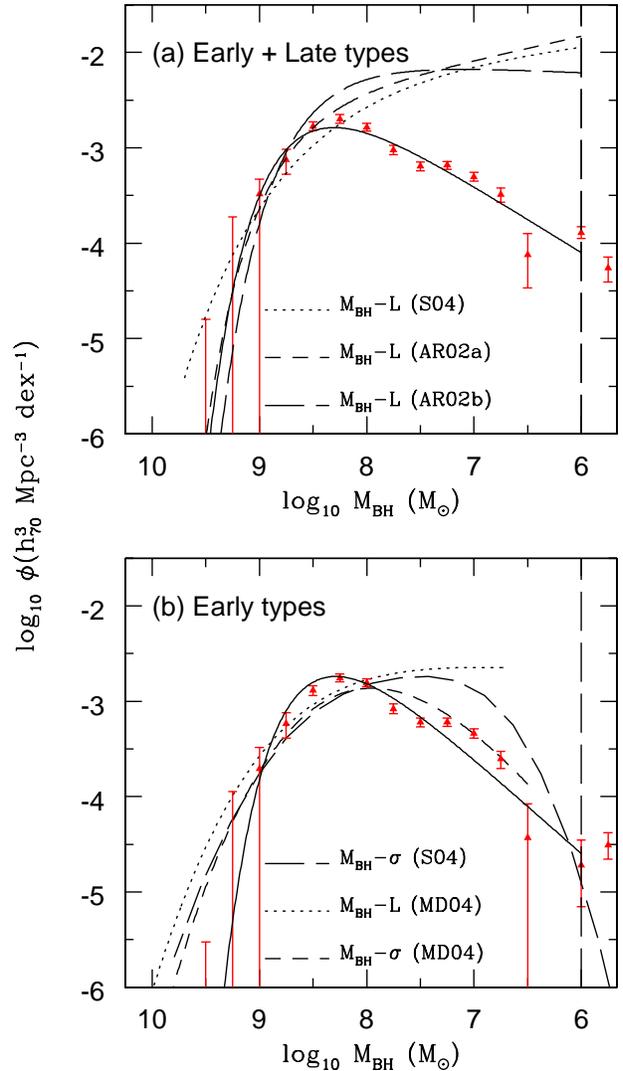}
\caption{ 
Our observed SMBH mass functions, along with our fits to Sample~3
(solid curves), overlaid with others estimates of the mass
function.  Shown in the top panel are curves from Shankar et al.\
(2004; S04) and Aller \& Richstone (2002):
(AR02)a is based on the Madgwick et al.\ (2002) luminosity function, and 
(AR02)b is based on the Marzke et al.\ (1994) luminosity function. 
Shown in the lower panel are curves from 
Shankar et al.\ (2004), and also McLure \& Dunlop (2004; MD04) 
who used two independent techniques (see Section~\ref{SecComp}). 
The ($M_{\rm bh}$--$\sigma$)-derived mass functions can be seen 
to turn down at low-masses, in agreement with our 
($M_{\rm bh}$--$n$)-derived mass function. 
We have corrected all curves to $H_0 = 70$ km s$^{-1}$ Mpc$^{-1}$, 
following the prescription given by Graham \& Driver (2007b). 
\label{mf2} 
}
\end{figure}

McLure \& Dunlop (2004) provided two estimates of the SMBH mass function (in
early-type galaxies) using two techniques.  At the low-mass end ($< 10^8
M_{\odot}$) their estimates do not agree; the mass function they obtained
using the $M_{\rm bh}$--$L$ relation gave noticeably higher SMBH number
densities than they obtained using the $M_{\rm bh}$--$\sigma$ relation
(Fig.\ref{mf2}, lower panel).  Both the method and data that we have used is
different to that used in McLure \& Dunlop (2004) and thus provides an
independent check on the SMBH mass function.  Our analysis, using the $M_{\rm
  bh}$--$n$ relation, shows a mass function which declines with decreasing
SMBH masses that are less than $10^8 M_{\odot}$, and therefore better matches
McLure \& Dunlop's mass function constructed using the $M_{\rm bh}$--$\sigma$
relation rather than the $M_{\rm bh}$--$L$ relation.

In plotting the mass functions from McLure \& Dunlop (2004) in Fig.\ref{mf2}, 
we have multiplied their ($M_{\rm bh}$--$L$)-derived masses by 1.70.  This
increase stems from the conversion of their $R$-band $M_{\rm bh}$--$L$
relation to the $K$-band. Starting from equation~6 in McLure \& Dunlop
(2002)\footnote{Transformation of 
equation~5 in McLure \& Dunlop (2002) would be considerably more
difficult because it was constructed from an inactive plus active
galaxy sample, a fraction of which have distances that depend on the
Hubble constant and also on the adopted cosmology given that some
of the active galaxies have redshifts which extend out to $\sim
0.5$.}
, one has 
\begin{eqnarray}
\log (M_{\rm bh}/M_{\odot}) = -0.5M_R - 2.91 \nonumber \\
 =  -0.5[M_K + 2.7 - M_{K,\odot} + 3.28 - 5\log(70/50)] -2.91 \nonumber \\
 = 1.25\log (L_K/L_{K,\odot}) - 5.53
\label{MD04}
\end{eqnarray}
Following McLure \& Dunlop (2002), we have assumed an average $R-K$
colour of 2.7.  We have used $M_{K,\odot}=3.28$ mag (Binney \&
Merrifield 1998).
McLure \& Dunlop (2002) used the SMBH masses derived using Tonry et
al.'s (2001) surface brightness fluctuation distances --- which are
independent of the Hubble constant.  However they did not use 
these $h$-independent distances in deriving the absolute magnitude of the bulges, but
used redshift derived distances and $H_0 = 50$ km s$^{-1}$ Mpc$^{-1}$.  
Equation~\ref{MD04} gives SMBH masses that are 0.23 dex larger
than equation~1 in McLure \& Dunlop (2004), and hence the factor of
1.70 (see Graham \& Driver 2007b for more details). 

In the lower panel of Fig.\ref{mf2}, one can see that our analysis 
suggests there may be a greater (up to a factor of $\sim$2) number 
density of SMBHs with masses $\log(M_{\rm bh}/M_{\odot}) \sim 
8.5$ than reported in McLure \& Dunlop (2004).  At smaller masses, $8 >
\log(M_{\rm bh}/M_{\odot}) > 6$, our mass functions roughly display
the same decline with decreasing SMBH mass as McLure \& Dunlop's
analysis based on the $M_{\rm bh}$--$\sigma$ relation.  Such a declining
SMBH mass function (for early-type galaxies) 
is also in qualitative agreement with Shankar et al.'s
(2004) mass function derived using a bivariate velocity
dispersion distribution.   In passing we note that we have 
increased Shankar et al.'s SMBH masses by 14 per cent, 
after adjusting for the correct dependence on the Hubble constant (see Graham 
\& Driver 2007b).  On the other hand, our data disagree with
mass functions that rise monotonically (which includes asymptotically)
towards lower-mass SMBHs, such as the ($M_{\rm bh}$--$L$)-derived 
mass functions in McLure \& Dunlop (2002), Aller \& Richstone (2002) and 
Shankar et al.\ (2004). 
%

The upper panel of Fig.\ref{mf2} reveals a clear disagreement, at masses below
$\sim 10^8 M_{\odot}$, between our (total) mass function and that derived from
$M_{\rm bh}$--$L$ relations.  We are not aware of any ($M_{\rm
  bh}$--$\sigma$)-derived mass function for late-type galaxies, and therefore
we do not show any such curve in this panel.  Some of the mismatch from the
($M_{\rm bh}$--$L$)-derived SMBH masses may stem from the methods used to
assign bulge flux in lenticular and spiral galaxies (see Graham \& Driver
2007b and Graham 2007).

Over the past few years a number of papers 
have constructed the local SMBH mass function and estimated the local SMBH
mass density, $\rho_{\rm bh,0}$.  We have listed several of these in
Table~\ref{Tab_Comp}.  It can be seen that our estimates of $\rho_{\rm bh,0}$
agree particularly well with the estimate in Fukugita \& Peebles (2004), and
is consistent (within the 1-$\sigma$ error bounds) with several other recent
studies.  It is worth noting, however, that our estimates are a factor of
$\sim$2 greater than reported in Wyithe (2006) and Aller \& Richstone (2002).
Due to the declining number density with decreasing SMBH mass, 
our adopted integration across the complete mass spectrum of SMBHs, rather
than truncating at $10^6 M_{\odot}$, does not significantly increase our
estimate of $\rho_{\rm bh,0}$.

In an effort to try and understand why our SMBH mass densities may be slightly
higher than reported by some, Graham \& Driver (2007b) has examined two
representative studies; one which used the $M_{\rm bh}$--$\sigma$ relation
(Aller \& Richstone 2002) and the other the $M_{\rm bh}$--$L$ relation
(Shankar et al.\ 2004).  Graham \& Driver identified a number of corrections
which can be made and showed that the mass density from Aller \& Richstone
(2002) is more than a factor of $\sim2$ too low.  As with the analysis by
McLure \& Dunlop, which was a factor of 1.7 too low, the reason is because of
over-looked dependencies on the Hubble constant.  These have been corrected in
Figure~\ref{mf2}, but not incorporated into Table~\ref{Tab_Comp}.

\begin{table*}
 \centering
 \begin{minipage}{125mm}
\caption{Local SMBH mass density estimates. 
The difference between Samples~1 and 3 (see Section~\ref{ColBias}) 
in our study is that the latter 
excludes galaxies with $(u-r)_{\rm core} < 2.00$ mag, such as
blue pseudo-bulges. 
The factor $h^3_{70} = [H_0/(70$ km s$^{-1}$ Mpc$^{-1})]^3$ 
is appropriate for our study
because the $M_{\rm bh}$--$n$ relation is independent of the Hubble
constant.  The majority of the densities from other papers have been
transformed to $H_0=70$ km s$^{-1}$ Mpc$^{-1}$ using $h^2$ rather than
$h^3$, as indicated in each paper.  However, as discussed in Graham \& Driver
(2007b), this may not always be appropriate. 
\label{Tab_Comp}
}
\begin{tabular}{@{}lccc@{}}
\hline
Study &  $\rho_{\rm bh,0}$ (E/S0)             &  $\rho_{\rm bh,0}$ (Sp)            & $\rho_{\rm bh,0}$ (total) \\
      &  $h^2_{70}10^5 M_{\odot}$ Mpc$^{-3}$   & $h^2_{70}10^5 M_{\odot}$ Mpc$^{-3}$ & $h^2_{70}10^5 M_{\odot}$ Mpc$^{-3}$ \\
\hline
This study (Sample~1)                         & $(3.8\pm 1.3) h_{70}$  & $(1.1\pm 0.6) h_{70}$ & $(4.9\pm 1.9) h_{70}$ \\
This study (Sample~3)                         & $(3.5\pm 1.2) h_{70}$  & $(1.0\pm 0.5) h_{70}$ & $(4.4\pm 1.7) h_{70}$ \\
Wyithe (2006)                                 &       ...            &       ...           & $2.28\pm0.44$       \\ 
Fukugita \& Peebles (2004)\footnote{See their equation~75.}  &  $(3.4^{+3.4}_{-1.7})h^{-1}_{70}$  &  $(1.7^{+1.7}_{-0.8})h^{-1}_{70}$  &  $(5.1^{+3.8}_{-1.9})h^{-1}_{70}$  \\ 
Marconi et al.\ (2004)                        &       3.3            &      1.3            & $4.6^{+1.9}_{-1.4}$ \\
Shankar et al.\ (2004)\footnote{Based on their ($M_{\rm bh}$--$L$)-derived mass function, and in agreement with their ($M_{\rm bh}$--$\sigma$)-derived values.}                  & $3.1^{+0.9}_{-0.8}$  & $1.1^{+0.5}_{-0.5}$ & $4.2^{+1.1}_{-1.1}$ \\
McLure \& Dunlop (2004)                       & $2.8\pm0.4$          &      ...            &    ...              \\
Wyithe \& Loeb (2003)                         &       ...            &      ...            & $2.2^{+3.9}_{-1.4}$ \\
Aller \& Richstone (2002)\footnote{Taken from their Table~2.}               & $1.8\pm0.6$          & $0.6\pm0.5$         & $2.4\pm0.8$         \\
Yu \& Tremaine (2002)\footnote{Based on their ($M_{\rm bh}$--$\sigma$)-derived mass function.}                         & $2.0\pm0.2$        & $0.9\pm0.2$       & $2.9\pm0.4$       \\
Merritt \& Ferrarese (2001)\footnote{See also Ferrarese (2002).}                   &        ...           &      ...            &    $4.6h^{-1}_{70}$   \\
Salucci et al.\ (1999)                        &      6.2             &      2.0            &    8.2              \\
\hline
\end{tabular}
\end{minipage}
\end{table*}

\section{Potential biases in our data}
\label{Sec_Corr}

\subsection{A Linear $M_{\rm bh}$--$n$ relation \label{SecLin}}

While we believe the quadratic relation between $M_{\rm bh}$ and $n$
(equation~\ref{EqQuad}) is the optimal expression to use when predicting
$M_{\rm bh}$ from $n$, we have explored how $\rho_{\rm bh,0}$ would 
change if we adopted the linear relation presented in Graham \& Driver 
(2007a, their equation~2).  
Although a linear fit to the data in Figure~\ref{MGC-m-n} results in
predictions of larger SMBH masses at the high-$n$ end, over the range $\sim$2
$< n <$ $\sim$8 it actually predicts lower SMBH masses (see Figure~3 in Graham
\& Driver 2007a).  Because the bulk of the MGC data falls in this interval
(see Figure~\ref{FigPaul})\footnote{We {\it speculate} that the apparent
  shortage of S\'ersic indices greater than $\sim$8 in the MGC data 
  (compared to Figure~\ref{MGC-m-n}) may arise
  from GIM2D's difficulty in deriving such large values while producing a
  galaxy magnitude that is close to the value obtained using SExtractor (due 
  to the large wings of high-$n$ profiles). 
  Confirmation of this speculation would require simulations 
  beyond the scope of this paper.  We therefore offer it as nothing
  more than speculation, but do note that if correct, a variety of plausible
  stretches to the MGC values of $n$ that we tested (to recover the possible
  true distribution) 
  resulted in a $\sim$20 per cent increase to the value of $\rho_{\rm bh,0}$
  (i.e., smaller than our current 1-$\sigma$ uncertainties on this value).}, 
use of the linear $M_{\rm bh}$--$n$
relation results in a 34 per cent lower SMBH mass density.  This value is
consistent, i.e.\ within the 1-$\sigma$ uncertainty, with our estimate
$\rho_{\rm bh, 0} = (4.4\pm 1.7) \times 10^5 h^3_{70} M_{\odot}$ Mpc$^{-3}$.
%
%
However, as noted in Graham \& Driver (2007a), this linear relation
has an intrinsic scatter of 0.31 dex, and so the estimate of $\rho_{\rm bh, 0}$
should be increased by 29 per cent, to give a value of 
$3.8 \times 10^5 h^3_{70} M_{\odot}$ Mpc$^{-3}$.

We have additionally removed the 
three high-$n$ data points from Figure~\ref{MGC-m-n} and obtained 
a new linear relation with the remaining 24 data points.
%
%
Doing so results in a relation with zero intrinsic scatter 
and a mass density that is 43 per cent higher than obtained with
the quadratic relation.  Specifically, we obtain 
a value of $(6.3 \pm 2.5) \times 10^5 h^3_{70} M_{\odot}$ Mpc$^{-3}$, 
which is only marginally higher than our upper 1-$\sigma$ 
value of $6.1 \times 10^5 h^3_{70} M_{\odot}$ Mpc$^{-3}$. 
%
%
%
We also find 
$\rho_{\rm bh, early-type} = (4.0 \pm 1.2) \times 10^5 h^3_{70} M_{\odot}$ Mpc$^{-3}$
and
$\rho_{\rm bh, late-type} = (2.3 \pm 1.3) \times 10^5 h^3_{70} M_{\odot}$ Mpc$^{-3}$.

\begin{figure}
\includegraphics[scale=0.42]{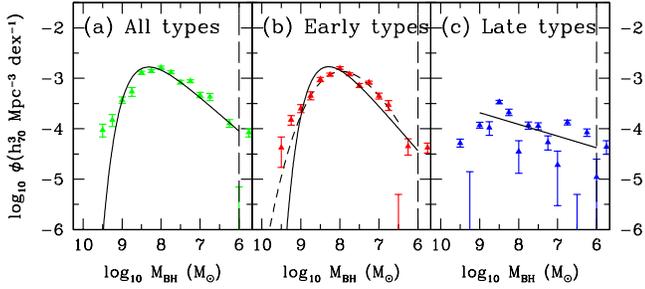}
\caption{Our (extreme) high-mass 
SMBH mass function constructed using the linear $M_{\rm bh}$--$n$
relation which excluded the three highest-$n$ galaxies (see
Section~\ref{SecLin}).
The solid curves are the optimal fits shown in Figure~\ref{Figfit}. 
The dashed curve in the middle panel is the ($M_{\rm bh}$--$L$)-derived
mass function given in McLure \& Dunlop (2004).
\label{Fig_lin}
}
\end{figure}

The associated SMBH mass functions are shown in Figure~\ref{Fig_lin} and
given in Table~\ref{TabLin}.  We caution that we do not believe these
are more accurate than those provided in Table~\ref{massfn_table_sb}, but that
they provide an (extreme) upper limit to the SMBH mass function and space
density.  As was noted in Section~2, the evidence for a curved $M_{\rm
 bh}$--$n$ relation is inherent in the low-n data. Excluding the five highest
$n$ galaxies results in a quadratic relation that is fully consistent with
equation~\ref{EqQuad}.  Removing the two galaxies with the highest SMBH
masses (one of which is suspected to be biased high: NGC~4486) produces the same
result.   The linear relation obtained after removing the three high $n$
galaxies is thus entirely driven by two galaxies (NGC~4486 and NGC~4649).

\begin{table}
\caption{
Supermassive black hole mass function data (corrected for Malmquist bias) for
the full, early- and late-type galaxy sample (Sample~3, see
Section~\ref{ColBias}) shown in Figure~\ref{Fig_lin}. 
The linear $M_{\rm bh}$--$n$ relation (Section~\ref{SecLin}), 
obtained after excluding the three galaxies
with the highest S\'ersic index, has been used here to provide an upper
estimate to the SMBH mass function. 
The uncertainties given are the upper and lower quartiles (i.e. $\pm$25 per cent) 
from extensive Monte Carlo realisation of the combined errors.} 
\label{TabLin}
\begin{tabular}{crrr} 
\hline
$\log_{10}$M$_{\rm bh}$ & \multicolumn{3}{c}{$\phi (10^{-4} h^3_{70}$ Mpc$^{-3}$ dex$^{-1})$} \\
$M_{\odot}$ & All galaxies & Early-type & Late-type \\ \hline
    5.00&    $0.12^{+0.12}_{-0.07}$ &    $0.06^{+0.04}_{-0.04}$ &   $0.08^{+0.12}_{-0.08}$ \\
    5.25&    $0.00^{+0.16}_{-0.00}$ &    $0.00^{+0.07}_{-0.00}$ &   $0.00^{+0.15}_{-0.00}$ \\
    5.50&    $1.13^{+0.17}_{-0.16}$ &    $0.00^{+0.08}_{-0.00}$ &   $1.26^{+0.14}_{-0.13}$ \\
    5.75&    $0.84^{+0.17}_{-0.15}$ &    $0.42^{+0.10}_{-0.09}$ &   $0.44^{+0.14}_{-0.13}$ \\
    6.00&    $0.00^{+0.19}_{-0.00}$ &    $0.00^{+0.13}_{-0.00}$ &   $0.11^{+0.14}_{-0.13}$ \\
    6.25&    $1.28^{+0.23}_{-0.21}$ &    $0.45^{+0.18}_{-0.15}$ &   $0.84^{+0.15}_{-0.14}$ \\
    6.50&    $0.00^{+0.39}_{-0.00}$ &    $0.00^{+0.42}_{-0.00}$ &   $0.00^{+0.15}_{-0.00}$ \\
    6.75&   $ 4.33^{+0.67}_{-0.70}$ &   $ 2.96^{+0.64}_{-0.67}$ &   $1.34^{+0.15}_{-0.15}$ \\
    7.00&   $ 4.68^{+0.58}_{-0.57}$ &   $ 4.48^{+0.57}_{-0.59}$ &   $0.19^{+0.17}_{-0.16}$ \\
    7.25&   $ 8.76^{+0.61}_{-0.60}$ &   $ 8.22^{+0.55}_{-0.55}$ &   $0.54^{+0.18}_{-0.16}$ \\
    7.50&   $ 8.37^{+0.75}_{-0.68}$ &   $ 7.21^{+0.69}_{-0.65}$ &   $1.13^{+0.18}_{-0.18}$ \\
    7.75&   $13.20^{+0.96}_{-0.88}$ &   $12.03^{+0.87}_{-0.83}$ &  $ 1.17^{+0.19}_{-0.19}$ \\
    8.00&   $16.23^{+1.14}_{-1.01}$ &   $15.88^{+1.01}_{-0.89}$ &  $ 0.35^{+0.23}_{-0.22}$ \\
    8.25&   $13.93^{+0.96}_{-0.90}$ &   $11.85^{+0.87}_{-0.80}$ &  $ 2.11^{+0.32}_{-0.29}$ \\
    8.50&   $12.89^{+0.93}_{-0.98}$ &   $ 9.52^{+0.90}_{-0.94}$ &  $ 3.39^{+0.27}_{-0.27}$ \\
    8.75&   $ 5.45^{+1.11}_{-1.09}$ &   $ 4.48^{+0.89}_{-0.76}$ &  $ 1.05^{+0.32}_{-0.31}$ \\
    9.00&    $3.69^{+0.57}_{-0.50}$ &    $2.51^{+0.48}_{-0.43}$ &   $1.18^{+0.17}_{-0.15}$ \\
    9.25&    $1.50^{+0.41}_{-0.39}$ &    $1.50^{+0.33}_{-0.33}$ &   $0.00^{+0.14}_{-0.00}$ \\
    9.50&    $0.94^{+0.29}_{-0.25}$ &    $0.42^{+0.27}_{-0.25}$ &   $0.52^{+0.10}_{-0.09}$ \\ 
\hline
\end{tabular} 

\noindent
Note: number densities are scaled to per unit $\log$M$_{\rm bh}$ interval and not per $0.25\log$M$_{\rm bh}$ interval.
\end{table}

\subsection{Blue Spheroids and Pseudobulges \label{ColBias}}

\begin{figure}
\includegraphics[scale=0.42]{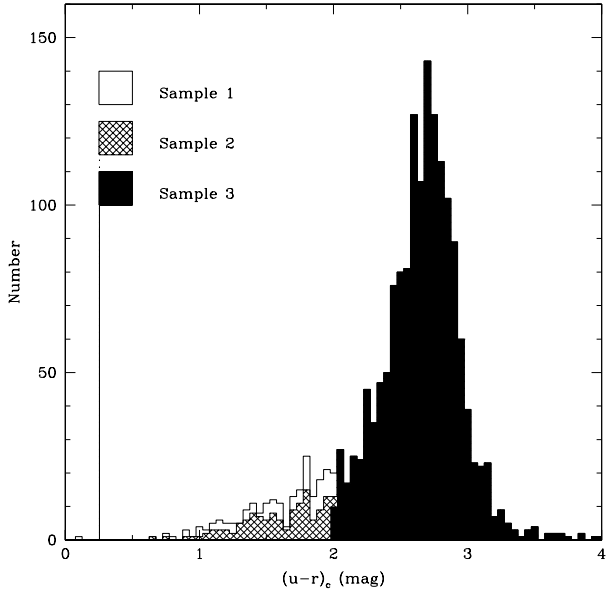}
\caption{
The $(u-r)_c$ core colour distribution of the three spheroid 
samples.  Only galaxies with $B/T>0.01$ and $M_B < -18$ mag
are shown.  A colour cut at $(u-r)_{\rm c}=2.00$ mag reflects the
central minimum observed in the bimodal distribution of core colours
shown in Liske et al.\ (2007, in prep.).
\label{Fig_ur}
}
\end{figure}

We divided each of the early-type (E/S0), late-type (Sp) and
(early + late)-type galaxy bins into three (colour) samples.  Sample~1 has had no
modifications to it, while Sample~2 excludes bulges (in disc galaxies) if the
($u-r$) core colour is bluer than 2.00 mag.  The core-colour is derived from
Sloan Digital Sky Survey point-spread-function magnitudes which have been
shown to maximise the colour bimodality, see Driver et al.\ (2006a).  This
colour-cut may help avoid possible pseudo-bulges (e.g., Erwin et al.\ 2003;
Kormendy \& Kennicutt 2004; Kormendy \& Fisher 2005) which may or may not have
SMBHs.
Sample~3 is further reduced to exclude all systems (i.e., bulges {\it and} 
elliptical galaxies) with $(u-r)_{\rm c} < 2.00$ mag 
(see Fig.\ref{Fig_ur}). 

It is not clear whether supermassive black holes exist in (every)
pseudo-bulge.  The tight correlation between the properties of dynamically hot
spheroids and their black hole mass has been tied to their joint formation
process.  As pseudo-bulges are believed to have formed via a distinctly
different formation process (i.e., from a redistribution of stars due to
perpendicular-to-the-disc resonances with a bar) it is plausible that
pseudo-bulges do not possess SMBHs.  Despite this conjecture, we note that
bars can fuel AGN growth (e.g., Ann \& Thakur 2005; Combes 2006; Ohta et al.\ 
2006).  Moreover, the Milky Way has a SMBH mass that appears consistent with
both the $M_{\rm bh}$--$n$ and $M_{\rm bh}$--$\sigma$ relations and probably
has a pseudo-bulge --- suggested from its rotation (e.g., Minniti et al.\ 
1992), flattening (e.g., Sharples, Walker \& Cropper 1990; Lopez-Corredoira et
al.\ 2005), presence of a bar (e.g., Lopez-Corredoira et al.\ 2006) and
peanut-shaped structure (e.g., Alard 2001, see also Debattista et al. 2006),
but see Zoccali et al.\ (2006) and Matteucci (2006). 
The situation is therefore ambiguous, which has resulted in our decision to
create three subsamples.

To unequivocally distinguish dynamically hot bulges from rotating
pseudo-bulges requires kinematical information.  In the absence of
this, an effective method is to use colour, with the pseudo-bulges
typically exhibiting blue colours comparable to a young (metal rich)
disc population and the genuine bulges exhibiting red colours
consistent with an old population.  We do however note that in some
cases the bulges are poorly resolved and the central colour is thus a
blend of the bulge and disc, which might result in some real bulges
erroneously falling blueward of our colour cut.
However the fact that the colour bimodality is clearly evident in our 
data suggests this is probably not a major issue, and 
we note that a positive correlation between bulge and disc colour 
is known to exist (e.g., Peletier \& Balcells 1996). 


Including blue spheroids 
with ($u-r)_c < 2.00$ mag, Sample~1 gives densities of 
$\rho_{\rm bh,early-type} = (3.8 \pm 1.3) \times 10^5 h_{70}^3 M_{\odot}$
Mpc$^{-3}$ and $\rho_{\rm bh,late-type} = (1.1 \pm 0.6) \times 10^5 h_{70}^3
M_{\odot}$ Mpc$^{-3}$ (Table~\ref{mgcrhos}).

\section{Summary}


We have used the log-quadratic 
relation between supermassive black hole mass
and the host spheroid's S\'ersic index $n$ (i.e.\ the 
$M_{\rm bh}$--$n$ relation, Graham \&
Driver 2007a) to predict the SMBH masses in galaxies from the
Millennium Galaxy Catalogue (MGC) for which a two-dimensional,
seeing-corrected, $R^{1/n}$-bulge plus exponential-disc decomposition
has been performed (Allen et al.\ 2006).

Applying the appropriate volume corrections, 
the number density of SMBHs with masses greater than $10^6 M_{\odot}$, 
and in spheroids more luminous than $M_B = -18$ mag, 
is $(2.3\pm0.1) \times 10^{-3} h_{70}^3$ Mpc$^{-3}$. 
%

We have constructed the local $(0.013 < z < 0.18, \bar{z}=0.12)$ 
SMBH mass function 
using direct estimates of the $M_{\rm bh}$ predictor-quantity $n$ 
(Fig.\ref{Figfit}).
Using every spheroid, the SMBH mass function appears
well represented by a 3-parameter Schechter function
having a logarithmic slope at the low-mass end of around two-thirds. 
Excluding the late-type galaxies, identified as systems with 
bulge-to-total ratios smaller than 0.4, the SMBH mass function 
was again well fit with a Schechter function, this time having a low-mass slope
of 1, and a turnover mass at $\sim 2\times 10^8 M_{\odot}$. 

Integrating down to a SMBH mass of $10^6 M_{\odot}$, the local mass 
density of SMBHs from early- and late-type galaxies combined is
$\rho_{\rm bh,0}= (4.4\pm 1.7) \times 10^5 h^3_{70} M_{\odot}$
Mpc$^{-3}$.  This value is slightly greater than, but still consistent
with, most values obtained via independent means over 
recent years (Table~\ref{Tab_Comp}, but see Graham \& Driver 2007b). 
%
Our mass density estimate can be split into 
$\rho_{\rm bh,early-type}= (3.5\pm 1.2) \times 10^5 h^3_{70} M_{\odot}$ Mpc$^{-3}$ 
and 
$\rho_{\rm bh,late-type}= (1.0\pm 0.5) \times 10^5 h^3_{70} M_{\odot}$ Mpc$^{-3}$, 
based upon a $B/T$ flux ratio cut at 0.4. 

In terms of the critical density of the universe, we obtain 
$\Omega_{\rm bh,total}=(3.2 \pm 1.2)\times 10^{-6} h_{70}$, or 
$\Omega_{\rm bh,early-type}=(2.5 \pm 0.9)\times 10^{-6} h_{70}$ and 
$\Omega_{\rm bh,late-type}=(0.7 \pm 0.4)\times 10^{-6} h_{70}$. 
%
%
Given that 4.5$h_{70}^{-2}$ per cent of the critical density is in the form of baryons
(Tegmark et al.\ 2006), 
the above figures imply $(0.007\pm0.003)h_{70}^3$ per cent of the 
total baryon content of the universe is locked up in SMBHs at the
centres of galaxies.


\section{acknowledgments}

A.G.\ would like to thank St Andrews University's School of Physics \&
Astronomy for its hospitality during a 3 week visit to work on this paper.
We are happy to thank Francesco Shankar for kindly supplying us with the data
from his ($M_{\rm bh}$--$\sigma$)-derived SMBH mass function shown in Shankar
et al.\ (2004, their Fig.7). We are additionally grateful to Francoise Combes
for her helpful comments about the Milky Way.
The Millennium Galaxy Catalogue consists of imaging data from the Isaac Newton
Telescope at the Spanish Observatorio del Roque de Los Muchachos of the
Instituto de Astrof\'{\i}sica de Canarias, and spectroscopic data from the
Anglo Australian Telescope, the ANU 2.3 m, the ESO New Technology Telescope,
the Telescopio Nazionale Galileo, and the Gemini Telescope.
This research has made use of the NASA/IPAC Extragalactic Database (NED) which
is operated by the Jet Propulsion Laboratory, California Institute of
Technology, under contract with the National Aeronautics and Space
Administration.
The survey has been supported through grants from the United Kingdom Particle
Physics and Astronomy Research Council, and the Australian Research Council
through Discovery Project Grant DP0451426.
The data and data products are publicly available from
http://www.eso.org/$\sim$jliske/mgc/ or on request from J.\ Liske or S.P.\ 
Driver.


\label{lastpage}
\end{document}